\documentclass[fleqn,usenatbib]{mnras}

\usepackage{newtxtext,newtxmath}
\usepackage[T1]{fontenc}
\usepackage{ae,aecompl}

\usepackage{graphicx}	% Including figure files
\usepackage{amsmath}	% Advanced maths commands

\usepackage[normalem]{ulem}
\usepackage[dvipsnames]{xcolor}

\newcommand{\Msun}{\ensuremath{{\rm M}_\odot}}
\title[Non-Halo structure lensing]{Non-Halo Structures and their Effects on Gravitational Lensing}

\author[T. R. G. Richardson et al.]{
{T. R. G. Richardson}$^{1,2,3,4}$\thanks{E-mail: thomas.richardson@obspm.fr (TR)},
{J. St\"ucker}$^{3}$,
{R. E. Angulo} $^{3,5}$,
{O. Hahn}$^{2,6,7}$
\\
$^{1}$ MAUCA — Master of Astrophysics, Universit\'e Côte d'Azur \& Observatoire de la C\^ote d'Azur, Parc Valrose, 06100 Nice (France)\\
$^2$ Laboratoire Lagrange, Universit\'e C\^ote d'Azur, Observatoire de la C\^ote d'Azur, CNRS, Blvd de l'Observatoire, CS 34229, 06304 Nice cedex 4, France.\\
$^3$ Donostia International Physics Centre (DIPC), Paseo Manuel de Lardizabal 4, 20018 Donostia-San Sebastian, Spain.\\
$^4$ Laboratoire Univers et Th\'eories (LUTh), Observatoire de Paris - PSL, CNRS, Universit\'e Paris Sciences Lettres, 5 Plc. Jules Janssen, 92190 Meudon, France \\
$^5$ IKERBASQUE, Basque Foundation for Science, E-48013, Bilbao, Spain.\\
$^6$ Department of Astrophysics, University of Vienna, T\"urkenschanzstraße 17, 1180 Vienna, Austria\\
$^7$ Department of Mathematics, University of Vienna, Oskar-Morgenstern-Platz 1, 1090 Vienna, Austria
}

\date{Accepted XXX. Received YYY; in original form ZZZ}

\pubyear{2020}

\begin{document}
\label{firstpage}
\pagerange{\pageref{firstpage}--\pageref{lastpage}}
\maketitle

\begin{abstract}
Anomalies in the flux-ratios of the images of quadruply-lensed quasars have been used to constrain the nature of dark matter. Assuming these lensing perturbations are caused by dark matter haloes, it is currently possible to constrain the mass of a hypothetical Warm Dark Matter (WDM) particle to be $m_\chi > 5.2$ keV. However, the assumption that perturbations are only caused by DM haloes might not be correct as other structures, such as filaments and pancakes, exist and make up a significant fraction of the mass in the universe, ranging between 5$\%$ -- 50$\%$ depending on the dark matter model. Using novel fragmentation-free simulations of 1 and 3keV WDM cosmologies we study these ``non-halo'' structures and estimate their impact on flux-ratio observations. We find that these structures display sharp density gradients with short correlation lengths, and can contribute more to the lensing signal than all haloes up to the half-mode mass combined, thus reducing the differences expected among WDM models. 
We estimate that non-halo structures can be the dominant cause of line-of-sight flux-ratio anomalies in very warm, but already excluded, $m_x \sim 1 \rm{keV}$ scenarios. For colder cases $m_x \gtrsim 3 \rm{keV}$, we estimate that non-haloes can contribute about $5 - 10\%$ of the total flux-ratio signal.
\end{abstract}


%%%%%%%%%%%%%%%%%%%%%%%%%%%%%%%%%%%%%%%%%%%%%%%%%%

%%%%%%%%%%%%%%%%% BODY OF PAPER %%%%%%%%%%%%%%%%%%

\section{Introduction}

One of the biggest puzzles of contemporary cosmology and particle physics is the nature of the dark matter (DM). DM dominates in mass by about five to one over ordinary matter \citep{Planck2018}, it is required to successfully reproduce many observations of the universe \citep[e.g.][]{Tegmark2004, de_Blok2008, Markevitch2004}, and it is an essential ingredient in cosmological $N$-body simulations that successfully predict the structure of the universe \cite[e.g.][for reviews]{Kuhlen2012,Frenk2012}.

Despite the cosmological evidence, there is no sign of a potential DM particle at at the Large Hadron Collider, nor at direct detection experiments (e.g. LUX, \citealt{LUX2018}, XENON1T \citealt{Xenon1T_2018}). This has made traditional candidates for DM increasingly less popular, while others such as axions \citep[e.g.][for reviews]{Sikivie2008,Marsh2016} or primordial black holes \citep[e.g.][for a review]{Carr2020} are enjoying renewed interest. 

Various competing DM models predict different features on cosmological scales, which opens up the opportunity to distinguish them observationally. For instance, sterile neutrino warm dark matter (WDM), with masses $\gtrsim 3\,{\text{keV}}$ \citep[e.g.][for a review]{Boyarsky2019}, or ultra-light axion-like particles with masses of $\sim10^{-20}\,{\text{eV}}$ forming a condensate of `fuzzy' DM \citep[FDM, e.g.][for a review]{Niemeyer2020}, are in agreement with all large-scale structure data but predict smooth rather than clumpy cosmic structure below a particle-mass-dependent scale. These differences are expected to affect various cosmological observables thus it becomes possible to constrain DM candidates and their properties.

There are currently four main venues to constrain DM with astrophysical observations: i) The number and properties of Milky Way satellites;   although these are found to be severely affected by astrophysical processes such as gas cooling and supernova explosions \citep[e.g.][]{Dekel1986,Ogiya2011,Pontzen2012,Zolotov2012,Arraki2014}, these galaxies are still sensitive to the amount of primordial small scale structure. ii) The amplitude of small-scale clustering of gas as measured by the Lyman-$\alpha$ forest. This method has put strong constraints on both WDM and FDM models down to scales that are now quite degenerate with astrophysical processes \citep[e.g.][]{Narayanan2000, Viel2013, Irsic2017,Kobayashi2017}. iii)  Perturbations to star density in stellar streams \citep[e.g.][]{Yoon2011, Banik2018,Banik2019,Hermans2020} have provided constraints on the population of sub-haloes around the MW and conversely the model of DM needed to produce this population. iv) Perturbations in  strong gravitational lensing which are becoming increasingly competitive in constraining the nature of DM thanks to recent advances in modelling \citep{Inoue2015a, Vegetti2018, Gilman2019, Hsueh2019, Gilman2020}. 
Recently, efforts have been made by the community to combine these various probes to reach more stringent constraints on the cosmological parameters \citep[see e.g.][]{Enzi2020, Nadler2021}

In this paper, we will focus on constraints obtained with observations of light fluxes of strongly-lensed quasars. The images of multiply-lensed high-redshift quasars originate from different light paths and have potentially encountered different structures which produce secondary lensing effects. This leads to anomalies in the flux-ratios of the images which cannot be explained by the main lens alone. These  ``anomalies'' are found to be sensitive to even very small DM structures, and can therefore be used to constrain the warmth of DM. Recent analyses of quadruply-lensed quasars have found that DM has to be colder than a thermal relic mass of $m_\chi < 5.2$ keV to explain the measured perturbations in the lensing signal \citep{Gilman2019, Hsueh2019, Gilman2020}.

In recent studies the amount of perturbation is directly linked to the abundance and concentration of DM haloes, implicitly neglecting any density fluctuation outside of haloes \citep{Gilman2019, Hsueh2019, Gilman2020}. However, FDM or WDM cosmologies are not completely devoid of small-scale structure outside haloes. Since haloes form by triaxial collapse \citep{Zeldovich1970,Shandarin1989}, only partially collapsed non-halo structures must exist: one-dimensionally collapsed `pancakes' and two-dimensionally collapsed `filaments'. Pancakes and filaments typically have lower densities than haloes, which is often the primary motivation for neglecting them. However, early and modern DM simulations have shown that these structures contain high-contrast caustics \citep{Buchert1989,Shandarin1989,Angulo2013} which create sharp high-density edges in the density field outside haloes. In CDM, the same structures exist, but are typically fragmented into even smaller substructures \citep[e.g.][]{Bond1996}.

As the precision of observations is increasing, it is important to review all the underlying assumptions in data analyses. Specifically, here we address the question of whether it is correct to assume that in a non-cold DM universe the only source of significant density fluctuations are collapsed haloes. In other words, can filaments and pancakes in a warm DM universe cause flux-ratio anomalies comparable with those of low-mass haloes in a colder cosmology? 

Pioneering work \citep{Inoue2015b} using theoretical models of non-halo structures attempted to answer this question for the strong lensing system MG0414+0534. Here we are able to address this question in a broader context thanks to a new generation of cosmological simulations \citep{Hahn2016, Stucker2020}, which, for the first time, simulate nonlinear structure with high precision and devoid of artificial fragmentation. Using the simulated density fields, we will create mock strong lensing-observations mimicking a quadruply-lensed high-$z$ quasar. First, we will study an idealized case where we align a WDM-filament with the lens geometry and show that filament could cause a relevant perturbation. Afterwards, we will create more realistic mocks of random projections of all (non-halo) structures in such WDM simulations to estimate their contribution to the total number of lensing perturbations. We will show that non-halo density fluctuation can indeed cause significant lensing anomalies, comparable in amplitude to the joint effect of all haloes below the half-mode mass of the corresponding WDM cosmology. 

The paper is organised as follows: in \S\ref{sec:wdm} we present our $N$-body followed by our gravitational lensing simulations in \S\ref{sec:lensing}. In \S\ref{sec:struct} we study a single filament extracted from our simulations. In \S\ref{sec:FR} we create a set of mock strong lensing observations and study the statistics of these measurements. Finally, in \S\ref{sec:conclusion} we discuss our results and present our conclusions.

\section{Simulations of WDM structure formation}
\label{sec:wdm}

In this section we describe our WDM numerical simulations. We first focus on the differences between CDM and WDM initial conditions, and then discuss our simulation technique. We refer to \cite{Stucker2020} and \cite{Stuecker2021} for specifics on our set of simulations.

\begin{table}
    \centering
    \caption{Parameters used in the simulations and throughout this work. The last line indicates the fraction of mass which was found to be outside of haloes at $z=0$.}
    \vspace{0.1cm}
    \begin{tabular}{cll}
    \hline
         Parameter  & 1 keV Sim. & 3 keV Sim.\\
    \hline
         $h$ & 0.679 & $-$\\
         $\Omega_{\text{m}}$ & $0.3051$ & $-$\\
         $\Omega_{\Lambda} $ & $0.6949$ & $-$\\
         $\Omega_{\text{K}}$ & $0$ & $-$\\
         $\sigma_8$ & $0.8154$ & $-$\\
         $M_\text{hm}$ & $2.5\cdot10^{10}h^{-1}\Msun$ & $5.7\cdot10^{8}h^{-1}\Msun$\\
         $L_\text{box}$ & $20h^{-1}$Mpc & $-$\\
         $N_\text{tracer} $ & $512^3$ &$-$\\
         $m_\text{tracer}$ & $5.0\cdot10^6h^{-1}\Msun$ & $-$\\
         $f_{\rm{non-halo}}$  & 45.7\% & 34.8\%  \\
         \hline
    \end{tabular}
    \label{tab:parameters}
\end{table}

\subsection{Initial conditions and simulation set-up}

The main difference between CDM and WDM is that the thermal velocities of the latter led it to free-stream out of small-scale perturbations in the early Universe, effectively suppressing their growth. As the universe expands, these initial velocities decay adiabatically and gravitationally-induced velocities grow. Hence, a very good approximation is to consider WDM as a cold system with a UV-truncated perturbation spectrum.  

To compute these initial fluctuation spectra for our WDM simulations, we use the parameterisation of the WDM transfer function by \cite{Bode2001} \citep[but see also][for an alternative parameterisation]{Viel2005}, where the matter density power spectrum for WDM is a low-pass filtered version of the CDM spectrum:

\begin{equation}
    P_{\text{WDM}}(k) = \left(1 + (\alpha k)^{-2} \right)^{-10} P_{\text{CDM}}(k)
\end{equation}

\noindent with 

\begin{equation}
\alpha = \frac{0.05}{h\,{\text{Mpc}}^{-1}}\left(\frac{\Omega_\chi}{.4}\right)^{0.15}\left(\frac{h}{.65}\right)^{1.3}\left(\frac{1 \text{keV}}{m_\chi}\right)^{1.15}\left(\frac{1.5}{g_\chi}\right)^{0.29}
\label{eq:bode}
\end{equation}

\noindent where $g_\chi = 1.5$, $m_\chi$ is the mass of the DM particle in units of keV, and $\Omega_{\chi}$ is the mean DM density in units of the critical density of the Universe. 

Here we will simulate cosmological structures in two WDM cases with $m_\chi=1\,{\text{keV}}$ and $3\,{\text{keV}}$ thermal relic WDM particles, as well as a CDM case. We will refer to these simulations using their respective thermal relic DM mass, i.e. the `1 keV simulation' or `3 keV simulation'. 

We note that the parameterization of the cut-off that we are using here is slightly different than the one that is most commonly used in the literature nowadays from \citet{Schneider2012}. This is so because these are simulations from a larger suite of simulations which explicitly triangulates the parameter space of possible cut-off functions as presented in \citet{Stuecker2021}. If our models are matched to the parameterization of \citet{Schneider2012} by matching at the half-mode mass, they correspond to slightly warmer cosmologies of $0.82$ keV and $2.6$ keV. However, this does not affect our conclusions, as we will mostly focus on qualitative considerations.

Our simulations consist of a cosmological volume of linear size $L_\text{box} = 20\,h^{-1}$Mpc. We generated both CDM and WDM initial conditions based on the same Gaussian noise field (implying that they share the same large-scale structure) using the {\sc Music} software\footnote{\href{https://www-n.oca.eu/ohahn/MUSIC/}{https://www-n.oca.eu/ohahn/MUSIC/}} \citep{Hahn2011,Hahn2013}. We use the cosmological parameters listed in Table~\ref{tab:parameters}. 

As we discussed above, here we assume that both CDM and WDM evolve as a collisionless fluid under their self-gravity and are thus governed by Vlasov-Poisson dynamics \citep[e.g.][]{Peebles1980}. In the case of CDM, the evolution can be followed by $N$-body techniques. However, WDM simulations are significantly more challenging numerically, as we will discuss next.

\subsection{Fragmentation-free WDM simulations}
\label{sec:sims}

\paragraph*{Artificial fragmentation.} \ $N$-body simulations have been very successful in predicting the non-linear evolution of cosmic structure from initial CDM perturbation spectra. However, already early simulations of a UV-truncated WDM perturbation spectrum showed that the same method performs significantly worse in this case, producing large amounts of spurious small-scale clumps instead of smooth structures \citep[e.g.][]{Wang2007} -- an effect which has been termed `artificial fragmentation'.

\paragraph*{Simulation method.} \ Recently, a new class of methods based on tessellations of the cold phase space distribution function \citep[cf.][]{Abel2012,Shandarin2012} has been developed by \cite{Hahn2013}, where the full three-dimensional hyper-surface of the cold distribution function is evolved. In this approach, the $N$-body particles serve as vertices of three-dimensional simplicial elements of the distribution function whose volume determines the DM density everywhere in space, without coarse graining. 

These methods have demonstrated to overcome the artificial fragmentation problem. However, in regions of strong mixing inside virialised structures, the number of vertices has to be increased to guarantee the tessellation still approximates well the distribution function. Adaptive refinement approaches to solve this problem have been proposed by \cite{Hahn2016} and by  \cite{Sousbie2016}. Since the number of required vertices can become exceedingly large \citep[cf.][due to phase and chaotic mixing]{Sousbie2016} inside of haloes (particularly so if high force resolution is used), most recently \cite{Stucker2020} have proposed a hybrid tessellation--$N$-body approach that resorts to the $N$-body method in regions where three-dimensional collapse has occurred based on a dynamical classification, and uses tessellations in the dynamically simpler voids, pancakes and filaments. 

This dynamical classification divides sheet tracing particles into four classes, voids, pancakes, filaments and haloes respectively corresponding to 0, 1, 2 and 3 collapsed axes, only the latter of the four having released N-body particles. This allows to trace and separate the different structures present in a simulation, a feature used in later sections.

\paragraph*{Resolution employed.} \ In our analysis, we use simulations based on this new approach proposed by \cite{Stucker2020}. Specifically, our simulations use $256^3$ particles to reconstruct the density field through interpolation of the phase space distribution function (where applicable), but additionally trace $512^3$ normal $N$-body particles which are used to reconstruct the density field where the interpolation fails (i.e. mostly in three-dimensionally collapsed structures) and can be used to infer other properties, such as the halo mass function. A more in-depth analysis of these simulations can be found in \citet{Stuecker2021}.

\paragraph*{Mass density field.} \ With similar interpolation techniques to those used to compute forces one can, in a post-processing step, recover a density field with much higher sampling than the original output \citep[e.g.][]{Abel2012,Hahn2016}. Other than generating detailed visualisations \citep[e.g.][]{Kaehler2012}, this feature also allows to recover small scale features that would not be visible from the initial tracer particles and to reduce the discreteness noise in lensing simulations \citep[e.g.][]{Angulo2014}. In later sections we will refer to the use of this technique as `resampling the density field'. Based on high-quality density fields generated in this way, we perform simulations of the strong gravitational lensing effect, which we describe next.

\section{Gravitational lensing simulations}
\label{sec:lensing}

In this section, we give a brief account of gravitational lensing theory, we then describe the technical details of our lensing simulations, and finish by discussing the measurements of simulated flux ratios of multiply-lensed sources.

\subsection{Theory of gravitational lensing}

We now recap the main equations of gravitational lensing and refer to one of the many reviews and textbooks for details \citep[e.g.][]{Schneider1992,Bartelmann2001,Dodelson2003}. 

Let $D_{\text{d}}$ and $D_{\text{s}}$ be the angular diameter distance from the observer to the deflector (i.e. the gravitational lens) and the source, respectively, and $D_{\text{ds}}$ that between deflector and source. In the `flat lens approximation' \citep{Bartelmann2001} the total ray deflection angle $\boldsymbol{\alpha}$ is related to the position of the image $\boldsymbol{\theta}$ and the position of the source $\boldsymbol{\beta}$ via:

\begin{equation}
    \boldsymbol{\beta} = \boldsymbol{\theta} - \boldsymbol{\alpha}.
\end{equation}

A prominent feature of this lens equation is that a single position in the source plane can map to several positions in the image plane, which originates multiple images in strong lensing.  

The deflection angle, $\boldsymbol{\alpha}$,  is the gradient of the ``lensing potential'' $\boldsymbol{\alpha} = \boldsymbol{\nabla}\psi$ which is given by a two dimensional Poisson equation:

\begin{equation}
    \nabla^2\psi = 2\kappa.
    \label{eq:poisson}
\end{equation}

\noindent where $\kappa$ is the ``normalised surface density'' defined as:

\begin{equation}
    \kappa = \frac{\Sigma(\boldsymbol{\theta}) - \bar\Sigma}{\Sigma_{\text{cr}}}
    \label{eq:def_kappa}
\end{equation}

\noindent and $\Sigma(\boldsymbol{\theta})$ is the projected surface density,
$\Sigma_\text{cr} = \frac{c^2}{4\pi G}\frac{D_{\text{s}}}{D_{\text{d}} D_{\text{ds}}}$ is the critical surface density, $c$ is the speed of light, $G$ is the gravitational constant, and $\bar\Sigma = \int_{0}^{z_{\text{bg}}}\rho_{\text{m}} dz $ is the mean surface density. 

Finally, the distortion matrix $\mathbfss{A}$ is defined as the Jacobian of the mapping between the source and image planes:

\begin{equation}
    \mathbfss{A} = \left|\frac{\partial\boldsymbol\beta}{\partial\boldsymbol\theta}\right|
\quad\textrm{i.e.}\quad
    \mathsf{A}_{ij} = \left|\partial_j\beta_{i}\right| = \left|\delta_{ij} - \partial_i\partial_j\psi\right| 
\end{equation}

\noindent where $\partial_i$ is the partial derivative with respect to the $i$-th coordinate of $\boldsymbol{\theta}$, and $\delta_{ij}$ is the Kronecker delta symbol. We chose to clarify the indexing convention due to the presence of numerical indices, represented by upper indices, that appear in later sections.

The inverse of the determinant of this matrix defines the magnification
\begin{equation}
    \mu = \frac{1}{\det\left[\mathbfss{A}\right]}    
\end{equation}
the curves in the image plane where $\det\left[\mathbfss{A}\right] = 0$ are called critical curves and limit the different regions where image replications are formed. The projections of these curves onto the source plane are called caustics and separate in which parts of the source plane are multiply imaged (Note that at these curves the magnification becomes infinite, but, in practice, astrophysical sources have a finite extent and so infinite magnification is never achieved.)

\subsection{Multiply-lensed quasar simulations}
\label{sec:lens_sim}

We now discuss how to numerically obtain deflection angles and distortion matrices from a simulated density field. We start by defining a grid in the image plane where we compute the normalised surface density and convergence field $\kappa$. We then express the lensing potential as a convolution

\begin{equation}
    \psi = g * 2\kappa.
    \label{eq:convol}
\end{equation} 

\noindent where $g$  is the Green's function of the two-dimensional Laplace operator $ g(\boldsymbol\theta) := \frac{1}{2\pi}\ln(\| \boldsymbol{\theta} \|)$. We note that we use the regularised integration kernel of \cite{Hejlesen2013} for solving the 2D Poisson equation avoiding the problem created by the singularity $\boldsymbol{\theta} = \boldsymbol{0}$ (c.f. Appendix~(\ref{app:num_imp}) for details). 

The deflection angles, distortion matrices, and magnification  can be obtained by exploiting the differentiation property of convolution \footnote{The differentiation property of convolution is 
\begin{equation}
    \partial_i(u(\bmath{x})*v(\bmath{x})) = \partial_i u(\bmath{x})*v(\bmath{x}) = u(\bmath{x})* \partial_i v(\bmath{x})
    \label{eq:diff_prop}
\end{equation}

\noindent where $u(\bmath{x})$, $v(\bmath{x})$ are generic distributions and $\partial_i$ is the derivative with respect to the $i^{\text{th}}$ component of $\bmath{x} $, such that we express the different quantities with respect to analytical derivatives of the Green's function.
}, which gives:

\begin{eqnarray}
    \alpha_i &=& \partial_i g * 2\kappa \\
    \mathsf{A}_{ij} &=& \delta_{ij} - \partial_i\partial_j g * 2\kappa,
\end{eqnarray}
where $\delta_{ij}$ is the Kronecker delta symbol and $\mathbfss{A}$ defines the magnification. We have thus obtained all the quantities needed for our application.

\paragraph*{Force splitting.} \ To efficiently incorporate the small and large-scale environment of the lens, we split the lensing potential as a sum of long and short range contributions, $\psi = \psi_s + \psi_\ell$, where:

\begin{equation}
    \psi_\ell = (2\kappa * g) * h_\ell \qquad\textrm{,}\qquad\psi_{\text{s}} = (2\kappa * g) * h_{\text{s}},
\end{equation}

\noindent and 

\begin{equation}
    \hat{h}_\ell := \exp\left\{ - 8 \pi^2 k^2 \ell^2\right\} \qquad\textrm{,}\qquad \hat{h}_{\text{s}} := 1 - \hat{h}_\ell,
    \label{eq:spliting_kernels}
\end{equation}

\noindent where $\ell$ is the splitting length-scale and the hat denotes the Fourier transform.

Operationally, we compute $\psi_\ell$ on a mesh with periodic boundary conditions that covers the full volume, and $\psi_{\text{s}}$ on a much finer mesh with non-periodic boundary conditions that covers only a small region around the lens and where the large-scale solution is interpolated linearly onto it.

\subsection{Flux ratio measurements}
\label{sec:quasars}
We now describe how we simulate quadruply-lensed quasars. We will first describe these observations, then we will present the methods used to simulate these systems using the previously presented implementation to solve the lensing equation.

\begin{figure}
    \centering
    \includegraphics[width = \linewidth]{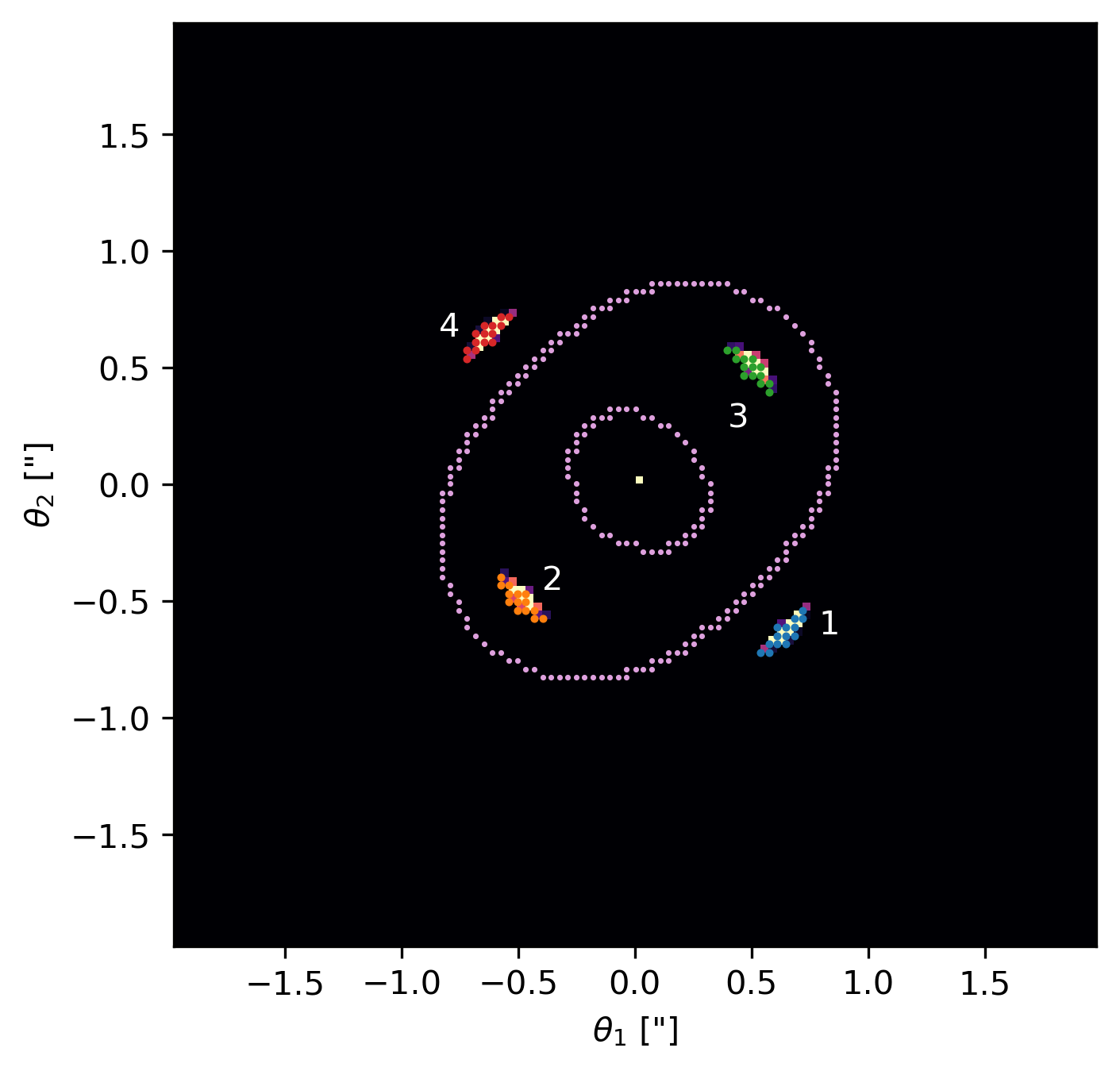}
    \caption{Distorted image produced by the reference lens model described in \S~\ref{sec:quasars}, showing the position of critical curves and detected multiple images. }
    \label{fig:quad}
\end{figure}

\paragraph*{Lens model.} \ We set up a simulation of a typical quadrupally lensed quasar system. The main deflector for these simulations is composed of a single projected elliptical NFW profile \citep{Golse2002} characterised by its mass $M_{200} = 4\times10^{13} \Msun$, its concentration $c_{200} = 8$, its eccentricity $\epsilon = 0.05$ and its orientation angle with respect to the main axes $\lambda = \pi/4$, the lens is placed at redshift $z_l = 0.29$ and the background source is placed at $z_{bg} = 1.71$ mimicking the redshift configuration of PG 1115+080 \citep[][]{Weymann1980, Chiba2005}. In this configuration the 4 images of a quasar placed exactly at the centre form at a radius $\theta_{\text{E}} \simeq 1$~\arcsec  as can be seen in Fig.~\ref{fig:quad}, in the lens plane this corresponds to $\theta_{\text{E}} D_{\text{d}} \simeq 3h^{-1}$kpc. In later paragraphs we refer to this distribution as our reference lens model.

We have checked the sensitivity of our results to the details of our adopted configuration by considering three different angular separations between the lens and the lensed quasar. These alternative configurations produced different absolute values for the flux-ratios, but all of them would give similar conclusions about the relative contributions of haloes and non-haloes -- which we are focusing on in this article. However, for simplicity we will restrict to the presentation of one configuration in this article.

\paragraph*{Source model.} \ We assume a point-like source and model the flux-ratios by the ratios of the magnifications at the image locations. (Note that Fig.~\ref{fig:quad} instead uses a disc of radius $0.01$\arcsec and constant intensity for purpose of visualization.)
We note that a point-like source is a simplifying assumption which might artificially increase the sensitivity to small structures, as we will discuss in Section \ref{sec:FR}.

\paragraph*{Image segmentation.} \ The alignment of the source and the deflector gives rise to five images, four located near the outer critical curve and a a fifth located close to the centre. The central image is heavily demagnified and often impossible to detect in typical observations, while the four outer images are strongly magnified and easily visible. In Fig.~\ref{fig:quad} we show the result of the same simulation where we have labeled the multiple images.

In each realisation, we measure the magnification at the location of each image $\theta$, where $\theta$ is inferred by solving numerically:
\begin{equation}
    ||\boldsymbol{\theta} - \boldsymbol{\alpha}(\boldsymbol{\theta}) - \boldsymbol{\beta_0}||^2 = 0
\end{equation}
\noindent using a two-dimensional root finding algorithm. The four different initial angles $\theta$ for solving this equation are found by using a projection of a slightly extended source into the image plane and using the center of each group (compare Figure~\ref{fig:quad}). We employ  \texttt{scipy.optimize.root}\footnote{\href{https://docs.scipy.org/doc/scipy/reference/generated/scipy.optimize.root.html}{\texttt{scipy.optimize.root} documentation}} \citep{Kelley1995, SciPy2020} readily available in {\sc Python}, using the Krylov approximation for the inverse Jacobian. Formally, as explained above, this equation admits 5 solutions. We therefore introduce an intermediate step where we use a cluster finding algorithm to locate the multiple images. We decided to use \texttt{scipy.ndimage.label}\footnote{\href{https://docs.scipy.org/doc/scipy/reference/generated/scipy.ndimage.label.html}{\texttt{scipy.ndimage.label} documentation}} \citep{Weaver1985, SciPy2020} readily available in {\sc Python}. Using as starting point the mean position of pixels belonging to a replicated image and repeating the minimisation for each image we ensure that we measure the magnification for all images and that the root finding algorithm doesn't converge twice, or more, to the same image.

\paragraph*{Flux ratios.} \ The main observable is the fluxes of the multiply imaged quasar. Under the hypothesis that the source has a small angular size with respect to the lens, we then estimate the flux of each image, $F_k$, as:
\begin{equation}
    F_k = \mu_k \, F_\text{int}  
\end{equation}
\noindent where $F_\text{int}$ is the intrinsic flux of the quasar and $\mu_k$ is the magnification measured at the position of the image. Since the intrinsic flux is unknown, it is not possible to recover the magnification of a single image. However, since all the images originate from the same background quasar, flux ratios remove the intrinsic contribution while retaining the information in the magnifications. These ratios are defined with respect to the brightest image produced by the model presented below, corresponding to image 4 in Fig.~\ref{fig:quad}. 
In this work we fix the lens model to the reference model described above. We then perturb this model's density field using different kinds of line-of-sight contributions. From this we are then able to investigate the the resulting flux-ratio anomalies caused by these line-of-sight contributions.

\paragraph*{Scale-filtered contributions} ~ However, in an actual observational scenario the base lens model is not known, but has to be fitted simultaneously. Therefore, if we wanted to mimic the observational situation, we would have to refit the lens model simultaneously on the perturbed lenses. For example, \citet{Gilman2020} do this by refitting the lens model so to keep the positions of the images fixed, while using the flux-ratios as the measured signal. We have tried to implement such a refitting procedure, but found in our case that the scatter induced by uncertainty in the lens model parameters is of the same order as the investigated signals, making reliable measurements impossible. This is so, since we use much smaller line-of-sights than \citet{Gilman2020}, since our density fields have to come from actual high resolution simulations.

Therefore, we choose another method to approximately distinguish between large-scale flux-ratio contributions that could be absorbed by the refitting of the lens model and small scale contributions that would cause measurable flux-ratio perturbations.
To this effect, we repurpose the force splitting kernels of Eq.~(\ref{eq:spliting_kernels}). Defining the large scale convergence field,
\begin{equation}
    \kappa_\ell = \kappa * h_\ell,
\end{equation}
and small scale convergence field,
\begin{equation}
    \kappa_\text{s} = \kappa * h_\text{s}.
\end{equation}
Here, we choose $\ell$ approximately equal to the Einstein radius, $\ell$= 0.85\arcsec, of our reference lens model. We show in Appendix \ref{app:scaledependence} that the precise value of this splitting scale does not significantly impact our main results.

\section{The Density field in WDM}
\label{sec:struct}

\begin{figure*}
    \centering
    \includegraphics[width = \linewidth]{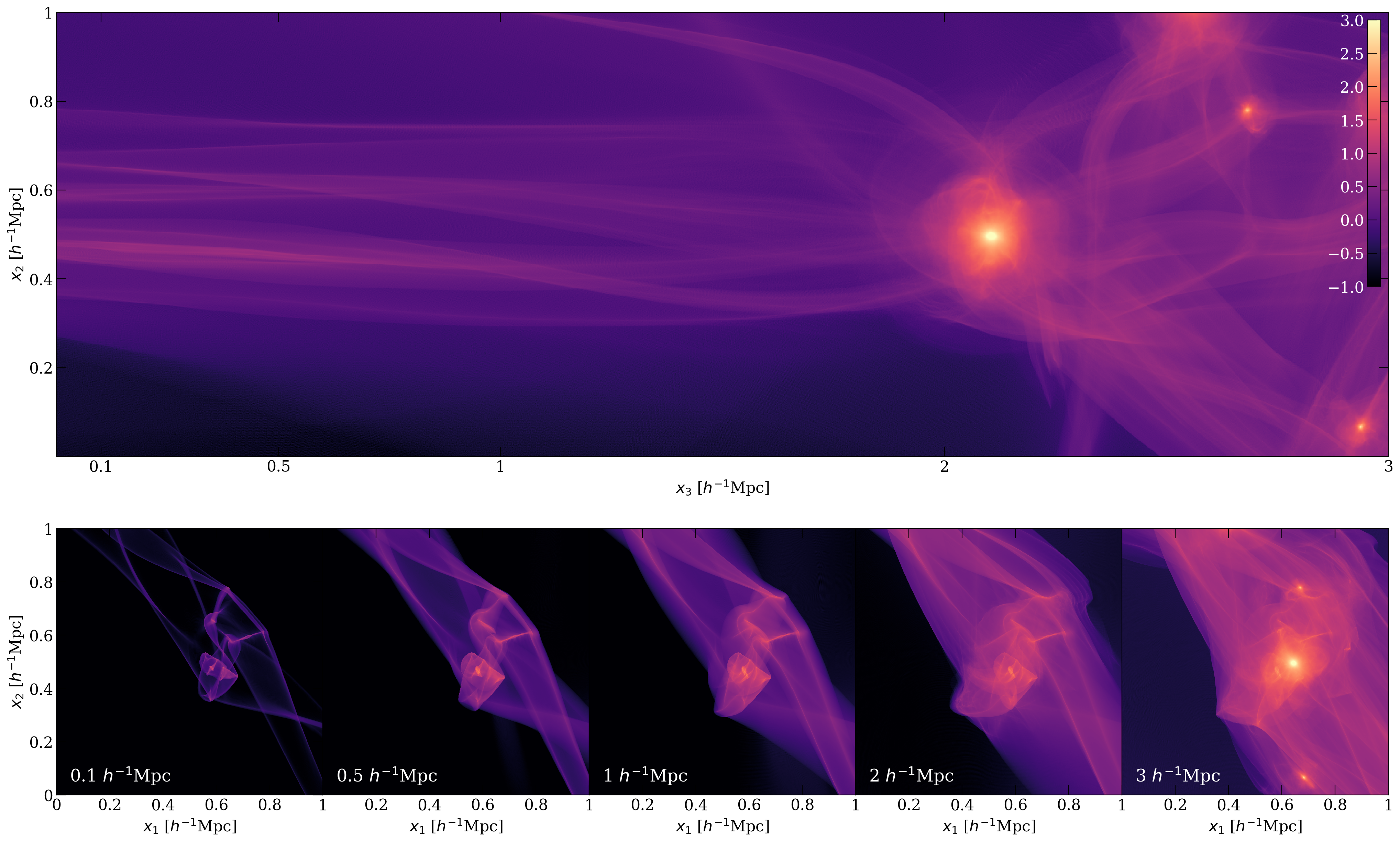}
    \caption{\textit{Top panel:} Column density of a filament from the 1 keV simulation in logarithmic units of $\rho_{\text{m}}h^{-1}$Mpc, seen side on. \textit{Bottom panels:} Same filament but projected along its axis for increasing projection depths. All panels are normalised to the same colour scale and the column density increases monotonically as the projection depth increases.  We can observe the diagonal structure in all projections, this structure is the pancake within which the filament is embedded. In the right most panel the haloes at the end of the filament are visible, it is these haloes that generate the high density tail of the corresponding black curve of Fig.~\ref{fig:dens}.}
    \label{fig:depths}
\end{figure*}

\begin{figure}
    \centering
    \includegraphics[width = 1.0 \linewidth]{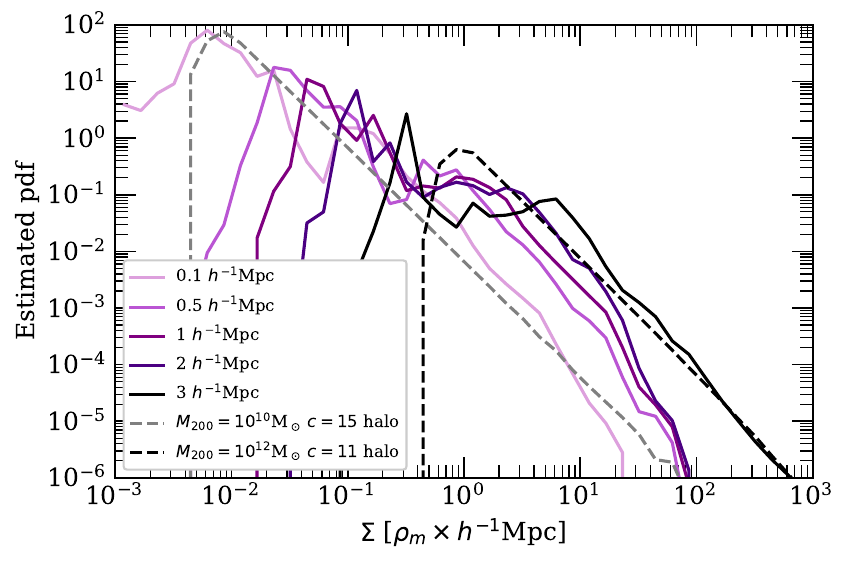}
    \caption{Column density distribution functions, $\phi(\Sigma)$, with $\Sigma$ in units of $\rho_{\text{m}} h^{-1}$Mpc for increasing projection depths.The high-density tail (likely caused by caustics) stops adding up coherently around $10^2 \rho_{\text{m}} h^{-1}$Mpc . The black curve is produced by the presence of high-mass haloes at the end of the filament as can be seen in Fig.~\ref{fig:depths}. The two dashed lines indicate the column density distributions of typical (CDM) NFW haloes with masses of $10^10 \Msun$ and $10^12 \Msun$. We see that the filament achieves peak column densities comparable with a $10^10 \Msun$ halo, while at the same time having much more total mass associated with intermediate density levels as well. }
    \label{fig:dens}
\end{figure}

In this section we will provide an initial exploration of the relative importance of halo versus non-halo structures in WDM simulations, and examine whether a single filament could create a lensing signal that is comparable with a halo at the WDM cutoff scale. 

\subsection{Haloes}
Haloes of WDM universes have already been investigated in numerous other studies \citep[eg.][]{Bode2001,Schneider2012,Angulo2013} and are reasonably well fitted by the two-parameter NFW profile \citep{Navarro1996,Lovell2014,Bose2016}. However, the abundance is heavily suppressed on small scales as a consequence of the dampening of the primordial power spectrum.

Using some simplifying assumptions, one can estimate the mass-scale below which the transfer function is suppressed by a factor of 2 with respect to the CDM counterpart. This characteristic ``half mode mass'' \citep{Viel2005,Schneider2012} is:

\begin{align}
    M_{\text{hm}} &= \frac{4\pi}{3}\rho_{\text{m}}\left(\frac{\lambda_{\text{hm}}}{2}\right)^3 \\
    \lambda_{\text{hm}} &=  2\pi\lambda_{\text{s}}^\text{eff} (2^{1/5} - 1)^{-1/2} \simeq 16.29\lambda^\text{eff}_{\text{s}}
\end{align}

\noindent with the effective free streaming scale $\lambda_{\text{s}}^\text{eff} = \alpha$ defined in Eq.~(\ref{eq:bode}). We provide the value of the half-mode mass for our 1~keV and 3~keV simulations in Table~(\ref{tab:parameters}).

Free streaming and the following suppression of small perturbations affects not only the abundance of small haloes, but also the density structure of haloes close to the half mode mass \citep{Bose2016, Ludlow2016}. Specifically, the concentration parameter of haloes of mass $M$ is modified as \citep{Bose2016}:

\begin{equation}
    \frac{c_{\text{WDM}}(M,z)}{c_{\text{CDM}}(M,z)} = (1+z)^{\beta(z)}\left(1 + 60\frac{M_{\text{hm}}}{M}\right)^{-0.17}
    \label{eq:c-m_relation}
\end{equation}
where $\beta(z) = 0.026z - 0.04$ and $z$ is the redshift. 

Here we will employ this model as a reference against which to compare non-halo structures both when it comes to their structure and the strong lensing signal they produce.

\subsection{The density structure of a filament}
\label{sec:densities}

A large fraction of the total mass of the universe is expected to reside beyond haloes in other structures like filaments, pancakes and voids. Estimates of the fraction of mass outside of haloes depend strongly on the free streaming scale of DM, but range between $5\%$ in very cold and $50\%$ in very warm scenarios \citep[e.g.][]{AnguloWhite2010,Buehlmann2019}. Thus, we would like to quantify if such uncollapsed mass could create an observable signature.

As an initial toy example, we selected a relatively long and straight (3.8 Mpc) filament from our 1 keV simulation. In addition, this filament does not contain any major halo or substructures. For this task we employ {\sc DisPerSE} \footnote{\href{http://www2.iap.fr/users/sousbie/web/html/indexd41d.html}{http://www2.iap.fr/users/sousbie/web/html/indexd41d.html}} \citep{Sousbie2011} which is an automatic structure identificator based on Morse theory. As discussed in \S \ref{sec:lensing}, the main quantity relevant for lensing is the projected density. Thus, as a best-case scenario, we will project the mass distribution along the primary axis of the filament. This will inform us if there is at least a small chance that a filament could generate a significant perturbation to a lensing signal.

In Fig.~\ref{fig:depths} we show our selected filament. We display density projections orthogonal to the filament's primary axis (top panel) and along its primary axis (bottom panels) with varying projection depths from 0.1 $h^{-1}$Mpc to 3 $h^{-1}$Mpc. We use the sheet resampling technique to increase the effective mass resolution by a factor of $64^3$ to $m\approx20 h^{-1}\Msun$. This is possible because the filament is still dynamically simple enough to be reconstructed by the interpolation algorithms described in \S \ref{sec:wdm}. 

In the top panel it can be seen that the filament exhibits several caustics which were formed by shell-crossing events and collapse along the filament minor axis. In the bottom left panel we show the slice orthogonal to the filament's axis which reveals a very rich structure. There is, for instance, a coherent structure going from the top-left to the bottom-right which corresponds to a pancake that this filament is embedded in. It can also be seen that the filament's internal structure is very different from the typical almost-isotropic inner structure of haloes. Instead its density structure is governed by multiple density peaks, sharp caustic edges, and different overlapping streams. 

In Fig.~\ref{fig:dens} we provide the distribution of column densities for varying projection depths (matching those shown in the bottom panel of Fig.~\ref{fig:depths}). It becomes clear that  column densities in filaments can reach orders of $100 \rho_{\text{m}}\,h^{-1}$Mpc\footnote{The value of $\rho_{\text{m}} h^{-1}$Mpc is $8.33\cdot10^{10}h^2 \Msun\text{Mpc}^{-2}$ at redshift $z = 0$.}. In comparison the average surface density of massive haloes of e.g.  $M_{200} = 10^12 \Msun$ and $c_{200} = 11$ on a disc of radius $r=3h^{-1}$kpc, the pertinent length for our reference lens model, is of the order $4\cdot10^3\rho_{\text{m}} h^{-1}$Mpc. On the other hand low mass haloes of e.g. $M_{200} = 10^10\Msun$ and $c=15$ reach only an average surface density of the order $4\cdot10^2\rho_{\text{m}} h^{-1}$Mpc comparable to the filament. 

An aspect of the filament projections worth noting is that even in this case, where we purposely aligned the filament with the line of sight, the high density features that we see in very thin slices do not add up very coherently and, instead, they get smeared out in the thicker projections. Therefore, the high column density tail does not get enhanced much beyond the $0.5 h^{-1}$Mpc projection.

 The toy example analysed in this section shows that filaments can indeed have significant densities with steep gradients, where the density can change by one or two orders of magnitude in a very small region. These sharp steps could potentially cause perturbations of a small enough coherence scale to be relevant in the flux-ratio measurements. To quantitatively estimate their relevance, however, not only the column densities, but their spatial pattern needs to be considered. 

\subsection{A strong gravitational lens perturbed by a line of sight filament} 
\label{sec:Fil_lensing}

To estimate quantitatively the effect a WDM filament can have as a lensing perturber, we create a mock lensing observation using the setup discussed in \S~\ref{sec:quasars}. 

Our lensing simulations are performed on a grid of side length $L = 0.1h^{-1}$Mpc $\sim23$~\arcsec and $N = 1024$ grid points per dimension for the fine grid, $L = 1h^{-1}$Mpc $\sim230$~\arcsec and $N = 4096$ grid points per dimension for the coarse grid and the splitting length $\ell = 0.92$~\arcsec. 

\begin{figure*} 
    \centering
    \includegraphics[width = \linewidth]{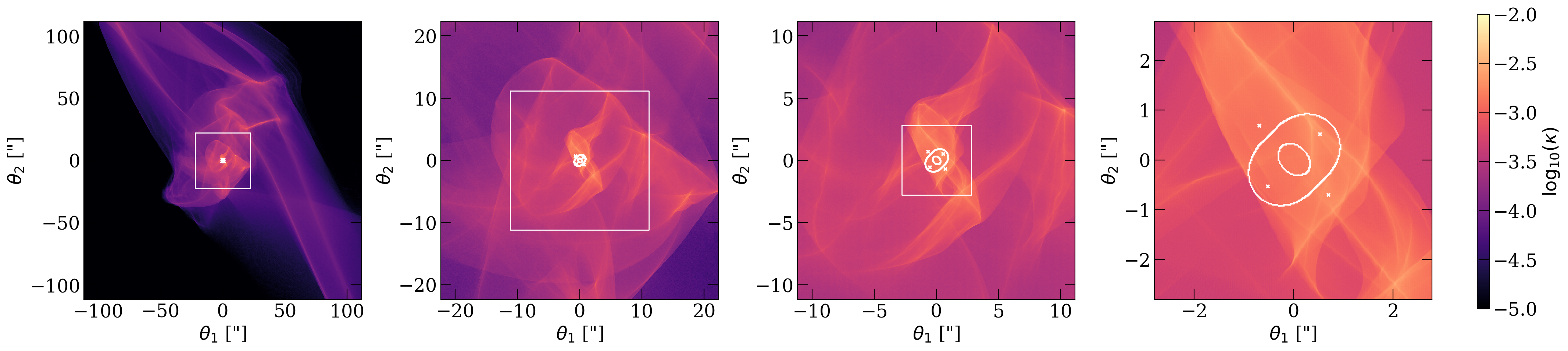}
    \caption{The filament used to perturb the lens in the first experiment. From left to right we show smaller and smaller scales. In each panel the white square represents the size of the panel to the right. In addition, we show the critical curves of the lens used in the first experiment along with the positions of the four replicated quasar images. Filaments can present significant density perturbations on scales smaller than the Einstein radius due to caustics.}
    \label{fig:scales}
\end{figure*}

We perturb our lensing simulations by adding the mass field corresponding to the $2h^{-1}$Mpc deep projection of the filament shown in Fig.~\ref{fig:depths}. We add the filament's projected density with varying offsets with respect to the main deflectors center -- on a path that takes it across the strongly lensed region. For each offset, we measure the magnification ratios of all quasar images. Each position is marked by its distance, $d$, from the centre of the strong lensing region. The sign of $d$ is such that when the perturbation is approaching the centre of the strong lensing region the sign is negative and inversely when it is moving away the sign is positive. For reference the radius at which the multiple images form is approximately $3$ to $4$ $h^{-1}$kpc as measured in the lens plane. In Fig.~(\ref{fig:scales}) we show successively enlarged projections around the central region of the filament along with the critical curves and positions of the replicated quasar images for scale. What can be observed is that the density field presents sharp caustic features at scales larger and smaller than the Einstein radius. 

\begin{figure}
    \centering
    \includegraphics[width = \linewidth]{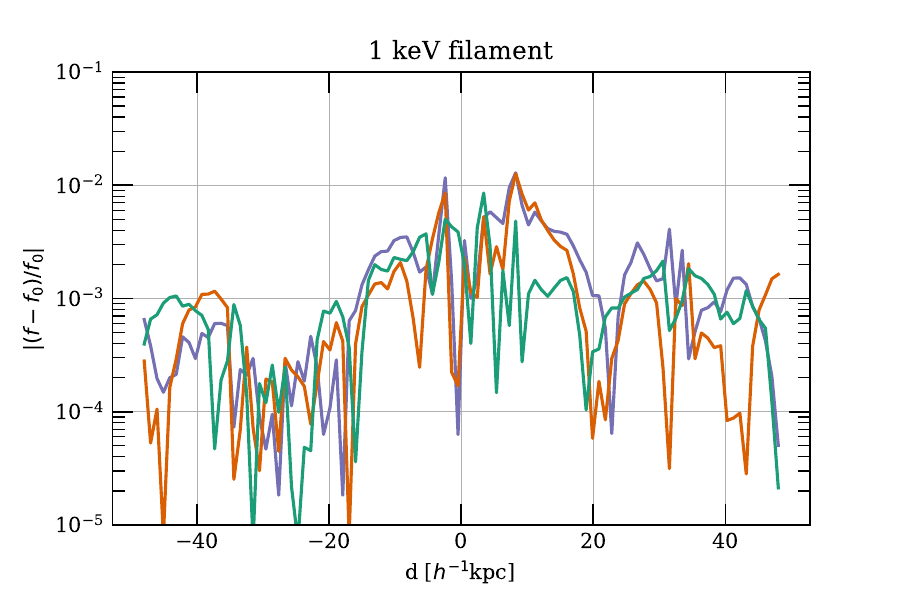}
    \includegraphics[width = \linewidth]{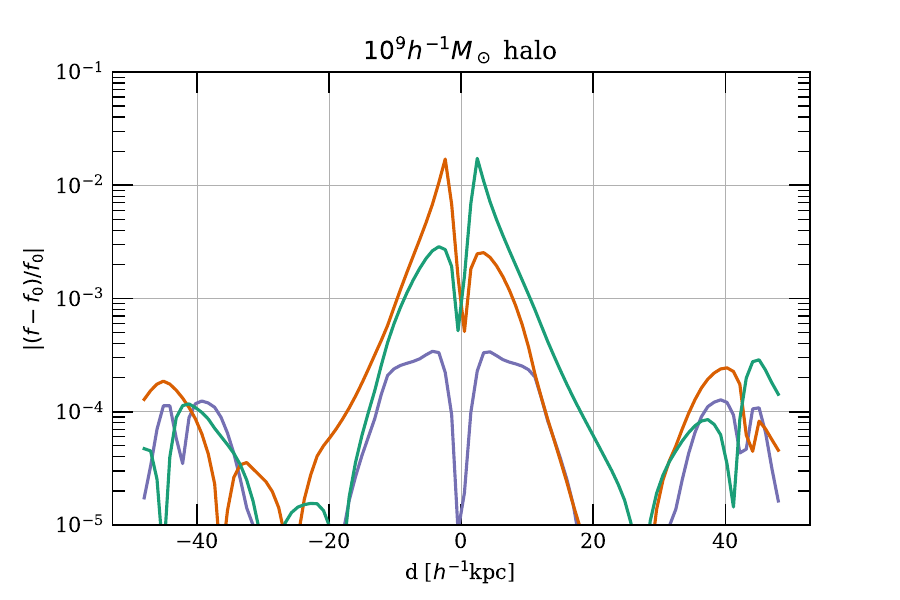}
    \caption{Relative change in flux ratio measurements in the presence of a perturbation, $f$ with respect to the unperturbed flux ratios $f_0$. The abscissas represent the offset of the perturbation with respect to the centre of lens plane in physical scale as measured in the lens plane. In this coordinate system, the multiple images of the quasar form at $~4 h^{-1}$kpc. Each line represents a different quasar image and are coloured according to the labelling of the images in Fig.~\ref{fig:quad}, the flux ratios are measure with respect to image 4 (red image). Top Panel: The lens is perturbed with a filament aligned with the line of sight for 2 $h^{-1}$Mpc. bottom panel: The lens is perturbed by a small 1 keV halo of mass $M_{200} = 10^9 \Msun$ and $c_{200}=5.82$. In both cases the perturbation follows the same path through the lens. One can observe that the filament is able to produce a considerable effect, similar to that produced by the halo.}
    \label{fig:perturb}
\end{figure}

For comparison, we repeat the same procedure, but using as a perturber a projected spherical NFW halo \citep{Golse2002} with mass $M_{200} = 1\times10^9 \Msun$ and concentration $c_{\text{WDM}} = 5.82$. This concentration corresponds to a typical halo at that mass, as given by the combined mass concentration relations from \citet{Ludlow2016} and \citet{Bose2016}. 

As described previously we remove the large scale contributions from the convergence perturbation field. The resulting images are not displaced by more than 5 m.a.s. allowing us to compare both scenarios.

In Fig.~\ref{fig:perturb}, we plot the difference of the flux ratios measured in presence of a perturbation, $f$, with respect to those in the case of the unperturbed lens, $f_0$. For the halo, the chosen path brings the centre of the perturbing halo close to two of the images, we see that the closer the structure is to the image the stronger the impact on the measurement.

This figure shows that the filament can cause perturbations of the same order of magnitude as the the $10^9 h^{-1} \Msun$ halo. The precise shape and amplitude, however, depends in a complicated manner on the alignment of the structures and can hardly be summarized in a simplified model. It will be the subject of the next section to test the impact of such non-halo structures in a  three-dimensional cosmological context.

\section{Flux ratio anomalies}
\label{sec:FR}

We have seen in the previous section that material outside haloes can in principle create flux-ratio perturbations comparable to small mass haloes. We will now perform a more quantitative study to see whether such lensing perturbations are likely or not. 

\begin{figure*}
    \centering
    \includegraphics[width = \linewidth]{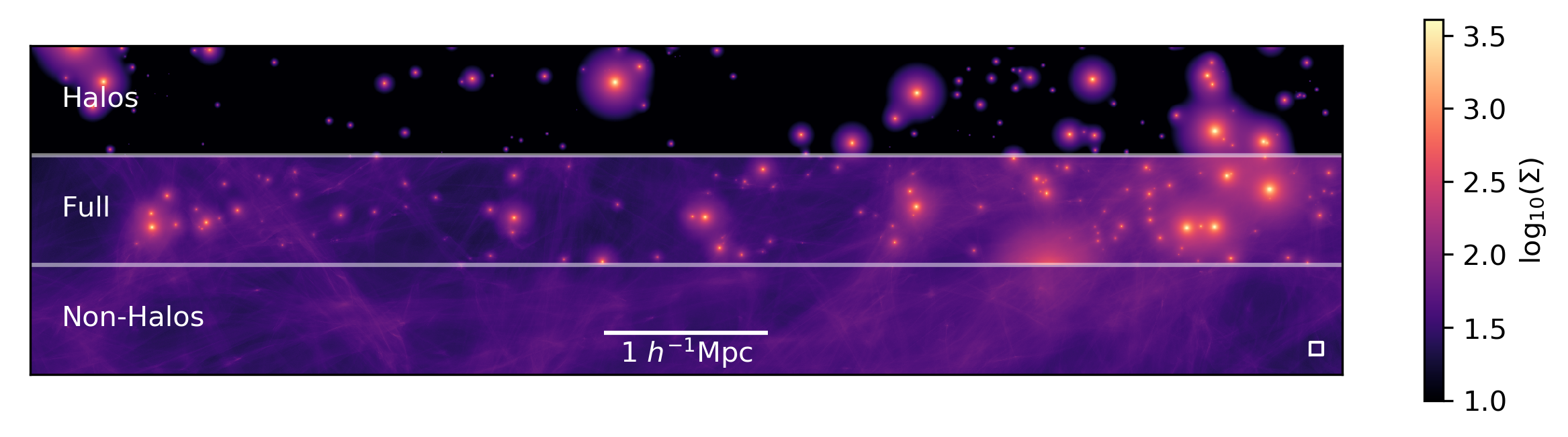}
    \caption{Exerts from the three large projections of the simulation volume. Top, we project only haloes as spherical NFW profiles. Bottom, we project only non-halo structures. Centre, we project both haloes and non-haloes simultaneously. This image has a width of $8h^{-1}$Mpc and a height of $2h^{-1}$Mpc. The small white square in the bottom right shows the size of the individual lines of sight to scale. }
    \label{fig:line_of_sight}
\end{figure*}

\begin{figure}
    \centering
    \includegraphics[width = 0.95\linewidth]{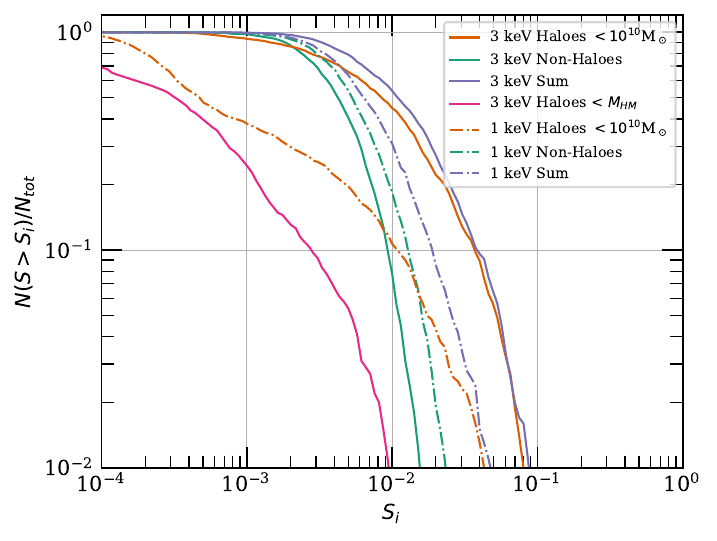}
    \includegraphics[width = 0.95\linewidth]{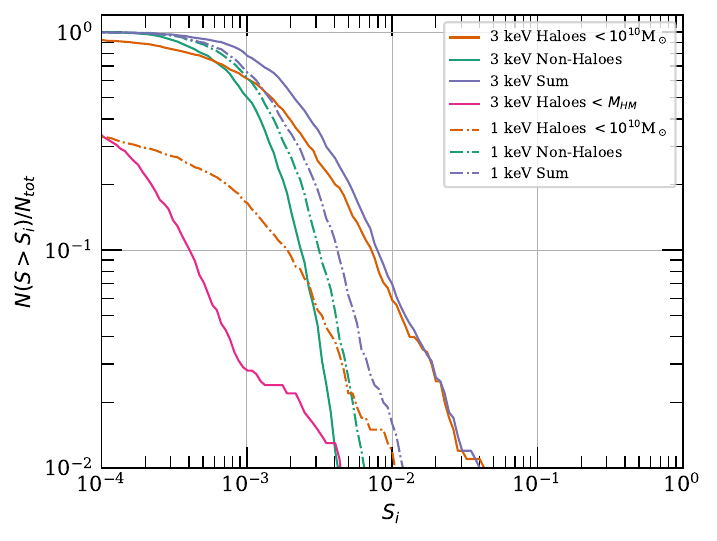}
    \caption{Cumulative histograms showing the fraction of summary statistics $S$ larger than $S_i$. The different colours show the statistics when modeling only haloes with masses $M_{200}<10^{10}\Msun$ (orange), only non-halo material (green) and both types simultaneously, \emph{Sum}, (blue). In the \textit{top panel} Using the full, i.e. large and small scale, contributions to the statistic and in the \textit{bottom panel} using only the small scale contribution. In very warm 1keV cosmologies the non-haloes constitute about half of the total flux-ratio signal. In colder cases, like 3keV, the relative contribution is much lower (around $5 - 10\%$). These estimates also hold when only considering the short-range part of the signal -- which cannot be absorbed by the fitting procedure in observational setups.}
    \label{fig:FR_cumulative}
\end{figure}

\subsection{Flux-ratio perturbations from random lines of sight}

We use our reference lens model, for which all the lensing quantities have analytical solutions, and perturb it according to mass distributions extracted from our WDM simulations along random lines of sight.

We consider deep projections ($8\times8\times80\,h^{-3}$Mpc$^3$) of the periodic simulation volume choosing a viewing angle that avoids replications of the same material in the projected volume and consider the projected volume only in the single lensing plane approximation. We note that given the size of our simulations $d= 20h^{-1}\text{Mpc}$ we cannot create projection depths comparable to the distance to typical strong lenses $\sim 1\text{Gpc}$. Therefore we do not attempt to quantify the absolute effect of the non-halo structures of the full line of sight. Instead we only consider the effects of the non-halo structures relative to the effect of haloes inferred from the same regions. We have checked for projection depths of $40\,h^{-3}$Mpc$^3$ and $160\,h^{-3}$Mpc$^3$ that the relative contributions of haloes and non-haloes stays roughly the same. We also see no reasons that the relative contributions should be affected by the simplified assumption of a single-plane approximation. In this restricted context one could consider our single lens as one lens within many in a multiplane formalism.

We construct density fields with two different kinds of projections:

\begin{itemize}
    \item The first projection considers \emph{only haloes}. To reduce numerical noise, we replace each simulated halo by a spherical NFW profile with a concentration that has been fitted to the profile. Beyond the virial radius, we then fade the profile with a cubic spline step function. We select objects with masses $M_{200}<10^{10}\Msun$ which are those expected not to host a detectable galaxy that can be directly inserted into a lens model \citep{Gilman2020}. For comparison, we also consider a case projecting only haloes below the expected $M_\text{hm}$ of the respective DM model.

    \item The second projection considers \emph{only non-halo} material. This is done by selecting mass elements that are not part of a halo according to the release criterion from \cite{Stucker2020} (c.f. their Eq.~17), and resampling them to a $\sim20\,\Msun$ mass resolution with the sheet interpolation method. Note that due to this high resolution resampling our lensing mocks do not suffer from discreteness effects like those inferred from traditional SPH / CIC density estimation techniques.
    
\end{itemize}
    
In Fig.~\ref{fig:line_of_sight} we show an example of the halo projection (without mass cut for this figure), the non-halo projection and the sum of both.

Within these large projections ($80 h^{-1}$Mpc deep) we then sample 1000 random lines of sight with a side length $L_\text{c}\simeq 160h^{-1}$kpc. The size of these cut-outs is chosen to be large compared to the radius at which the multiple images form ($\theta_{\text{E}}D_{\text{d}} = 3 h^{-1}$kpc with respect to the centre of the lens plane). The small square in the bottom right of Fig.~\ref{fig:line_of_sight} displays the size of these regions.

These cut-outs are then used to compute the gravitational lensing effect on our fine grids $(N = 1024)$ while the large scale contribution is calculated with the full large projection using $N=8182$ grid points per dimension and a splitting length $\ell = 1.84$~\arcsec. As such, the fine grid is calculated out to $50\,\theta_{\text{E}}$ while the coarse grid is calculated out to $2900\,\theta_{\text{E}}$ to capture the influence of the large scale environment.

Using the linearity of Eq.~(\ref{eq:poisson}) we then add this perturbation to the analytical reference lens model. With a simulated quasar placed at redshift $z_{bg} = 1.71$  and sky coordinates that compensate for the mean deflection generated by the large scale environment, which ensures the quasar is quadruply lensed, we measure the magnification of each image and calculate the three flux ratios, $f_i$.

For each line of sight, we can quantify the perturbations to the flux ratios using the following summary statistic:

\begin{equation}
    S := \sqrt{\sum_{i=1}^{3}(f_i - f_{\text{ref}(i)})^2},
\end{equation}

\noindent where $f_{\text{ref}(i)}$ is a flux ratio for image $i$ as measured with the reference unperturbed model, note this is the same statistic as used by \cite{Gilman2019}. The larger the value of $S$, the larger is the expected perturbation to the main lensed images. We remind the reader that in our case the absolute values of $S$ are about an order of magnitude smaller than of typical lenses, since we have a quite short line of sight of $80$Mpc$/h$. However, the relative comparison between the $S$-distribution of haloes and non-halo structures should be unaffected by this since both scale with the length of the line of sight.

In Fig.~\ref{fig:FR_cumulative} we show the cumulative distributions for $S$ from our $1000$ lines of sight. These distributions can be interpreted as the probability of observing a flux ratio anomaly higher than a given level. 
Lines of different colours correspond to different types of structures and dashed lines to the 1 keV simulation versus solid 3 keV lines. The top panel shows the flux-ratios obtained when using the full convergence field whereas the bottom panel shows the flux-ratios from the high-pass filtered convergence field as explained in Section \ref{sec:quasars}. Therefore, the top panel includes contributions that might be absorbed into the parameters of the lens model in observational studies, whereas the bottom panel represents short-range contributions that cannot be absorbed into the lens model. We investigate the flux-ratio anomalies as a function of scale further in Appendix \ref{app:scaledependence} where we find that the relative contributions of haloes and non-haloes remain similar on large and small scales.

First of all we see that haloes from cosmologies with different warmth (solid versus dashed orange lines in the last panel), indeed produce different statistics for the flux-ratio perturbations. This is the effect that is used to constrain the warmth of DM from flux-ratio observations \citep[eg.][]{Hsueh2019, Gilman2020}.

Second, we concentrate on the curves from the 1keV universe (dashed lines). We can see that in both -- the unfiltered and the filtered case (top versus bottom) -- non-haloes have a significant contribution to the flux-ratio signal. In the short-range case (bottom) the non-haloes even dominate over the contribution from all haloes with $M < 10^{10} h^{-1}\Msun$ at almost all perturbation levels. This shows that non-halo structures cannot be neglected when estimating the flux-ratios in very warm cosmologies -- like 1keV -- and that their contribution cannot be absorbed by the parameters of the lens (like the shear or a uniform mass-sheet).

Next, we consider the flux-ratio signals in the colder 3keV case (solid lines). First of all, we see that in a 3keV cosmology the non-haloes still have a larger contribution to the signal than all haloes below the half-mode mass combined ($M < M_\text{hm} = 5.7\cdot10^8 h^{-1}\Msun$). However, the dominant contribution to the halo signal comes from larger masses, so that the haloes with $M < 10^{10} \Msun$ have a much larger signal than the non-haloes -- both in the filtered and the unfiltered cases. When comparing the `sum'-line with the halo-line we note that neglecting the non-haloes in the 3keV cosmology causes only a minor underestimate of the flux-ratios by $5-10\%$. However, we note that our simplifying assumption of a point-like source might artificially increase the sensitivity to small haloes. Therefore, the relative contribution of filaments to the total signal might even increase a bit when considering an extended source.

When comparing the ``non-halo'' contribution in the WDM simulations, we see that it decreases slightly, by $20-30\%$, between the 1keV and 3keV models. Since the fraction of total mass outside of haloes,  $f_{\rm{non-halo}}$, is smaller in colder models, more mass is collapsed into small haloes. As mentioned in Table \ref{tab:parameters}, for the 1keV simulation we have $f_{\rm{non-halo}}=46\%$ and for the 3keV universe $f_{\rm{non-halo}}=35\%$. The ratio between these numbers is roughly consistent with the shift in $S$. In simple excursion set models, the fraction of mass outside haloes changes very slowly with the cut-off scale. Even for very cold DM models, such as a 100GeV neutralino,  about $5\%-20\%$ of mass is expected be to reside outside of haloes at $z=0$. This fraction increases significantly at higher redshifts  \citep[e.g.][]{AnguloWhite2010}. Therefore, we speculate that the ``non-halo'' contribution could shift from the $3 $keV case to slightly, but not significantly, smaller values for colder dark matter candidates.

These results suggest that non-halo material is not only dense enough to cause lensing perturbations, but that it commonly does so in WDM scenarios. The effect of non-haloes can be dominant over haloes in relatively warm cosmologies like 1keV thermal relic warm dark matter, which are, however, already excluded observationally. In colder cases (like 3keV) the absolute effect of non-haloes stays similar, but the relative effect decreases since the effect of haloes becomes stronger -- leading to relative contributions at the order of $5\%-10\%$. Therefore, the often made assumption that non-halo structures can be neglected for the modeling of flux-ratio perturbations, holds within certain limits. Constraints on cosmologies with $m_x > \rm{3keV}$ should be unaffected if they are robust to prediction biases of order $10\%$. However, we recommend careful investigations of the effect of non-haloes if a higher accuracy is needed to tell the difference between dark matter models.

\section{Conclusions}
\label{sec:conclusion}

The main question of this study was: \emph{can the material outside of haloes -- which resides for example in filaments or pancakes -- cause relevant effects in strong gravitational lenses?} In particular, we have focused on lensing systems where a quasar is quadruply lensed. In such systems the ratios of fluxes from the different images can be used to constrain line-of-sight structures and thereby the DM warmth \citep{Gilman2019,Hsueh2019,Gilman2020}. So far all studies have only modeled the effect of haloes, but not any other structures in the density field. However, for example the caustic structure of a filament in a WDM universe can create density variations on very small scales -- even when no small haloes are expected at these length scales. It is therefore important to understand the effect of non-halo structures qualitatively and quantitatively.

In a first qualitative part of this study we have shown the caustic structure and the sharp density edges that exist in projections of a warm DM filament without substructure. Further we have confirmed that such a filament creates a relevant perturbation when aligned with the line of sight, and found that its effect is comparable that of a halo of $10^9 \Msun$. This shows that, at least in principle, such a filament could cause significant effects in flux-ratio observations.

In a second -- more quantitative -- investigation we have created a large number of mock lensing observations from random projections of two state-of-the-art WDM simulations. From these measurements we have found that the flux-ratio perturbations created by non-halo structures can be larger than those caused by all haloes up to the half mode mass of the corresponding cosmology.

Because of the numerical requirements, we only simulated warm dark matter models here which are already excluded by current constraints. However, we argue that the relative importance of the non-halo structures (in comparison to haloes around the half-mode mass) becomes even larger for colder DM candidates (e.g. $m_{\chi} \sim 5$ keV). We speculate that the effect of non-halo structures is roughly proportional to the total fraction of mass outside of haloes, and therefore it should only exhibit a moderate decrease when considering colder models. On the other hand, perturbations from haloes around the half-mode mass decrease rapidly with decreasing  half-mode mass. Even in cold dark matter cosmologies a significant fraction of mass resides outside of haloes and might have an impact on flux-ratio lensing observations.

A precise quantification of the effect on recent constraints of the warmth of DM would require mock simulations mimicking details of the observations and analyses. However, in our simplified setup we found that non-haloes dominated the flux-ratio signal in very warm, but already excluded cosmologies. For colder cosmologies, we estimate that non-haloes can have a relative contribution of $5-10\%$ to the flux-ratio signal from line-of-sight objects. Neglecting this component may therefore underestimate the likelihood of observing certain flux-ratio anomalies and may lead to a bias in favour of colder dark matter candidates.
We conclude that the effect of non-haloes has to be included in observational studies if such a high prediction accuracy is needed to tell the difference between different dark matter models.

We have made a variety of simplifications when estimating the perturbations of flux-ratios. These include that we only modeled a single base-lens; we could only project over a relatively short line of sight ($80h^{-1}$Mpc); we worked in the single-lens approximation, we did not model baryonic effects; and we only used simulations of two quite warm DM models incompatible with current constraints. These simplifications prevent us from making an absolute estimate of the frequency of anomalies caused by non-halo material, however, we expect the relative contribution with respect to that of halos to be largely unaffected by our assumptions, and thus we argue for the robustness of our conclusions.

We conclude that a rigorous modelling of flux-ratio anomalies should include the effect of non-haloes. However, since the inclusion of these non-halo structures is quite difficult in practice, they may be neglected under some circumstances as an approximation. We estimate that they can be neglected for cosmologies with $m_x > 3\rm{keV}$ if a bias of $5-10\%$ in the prediction of the total flux-ratio signal is acceptable.

\section*{Acknowledgements}
We acknowledge insightful comments on the manuscript from Simona Vegetti, Carlo Giocoli,
Giulia Despali, Kiaki Taro Inoue, Simon Birrer and the anonymous referee. T.R. thanks the Observatoire de la C\^ote d’Azur, the DIPC, and the Erasmus+ program for making this joint research work possible. J.S., R.A. and O.H. acknowledge funding from the European Research Council (ERC) under the European Union's Horizon 2020 research and innovation programmes: Grant agreement No. 716151 (BACCO) for J.S. and R.A.; and grant agreement No.\ 679145 (COSMO-SIMS) for O.H. The authors thankfully acknowledge the computer resources at MareNostrumIV and technical support provided by the Barcelona Supercomputing Center (RES-AECT-2019-3-0015).

%The Acknowledgements section is not numbered. Here you can thank helpful
%colleagues, acknowledge funding agencies, telescopes and facilities used etc.

\section*{Data Availability}
The data underlying this article will be shared on reasonable request
to the corresponding author.

%%%%%%%%%%%%%%%%%%%%%%%%%%%%%%%%%%%%%%%%%%%%%%%%%%



\bibliographystyle{mnras}
\bibliography{bibliography} % if your bibtex file is called example.bib

%\begin{thebibliography}{99}

%\end{thebibliography}

%%%%%%%%%%%%%%%%%%%%%%%%%%%%%%%%%%%%%%%%%%%%%%%%%%

%%%%%%%%%%%%%%%%% APPENDICES %%%%%%%%%%%%%%%%%%%%%

\appendix

\section{Numerical implementations}
\label{app:num_imp}

In this section we discuss the specifics of the numerical implementations used to solve the lens equation.

\paragraph*{Computation of the lensing potential} \ As said in \S~(\ref{sec:lensing}) the gravitational lensing potential is defined by Eq.~(\ref{eq:poisson}), a poisson equation which can be solved by a convolution with the appropriate Green's function, see Eq.~(\ref{eq:convol}). The analytical Green's function for the two-dimensional Laplace operator presenting a singularity at the origin, which prevents convergence, we instead use the regularised integration kernels 
\begin{equation}
    g_m = -\frac{1}{2\pi}\left[\ln(\theta) - Q_m\left(\frac{\theta}{\epsilon}\right)\exp\left(-\frac{\theta^2}{2\epsilon^2}\right) + \frac{1}{2}E_1\left(\frac{\theta^2}{2\epsilon^2}\right)\right]
\end{equation}
proposed by \cite{Hejlesen2013}. Where $\epsilon$ is a smoothing parameter set to 1.5 times the grid spacing $\delta$, $Q_m$ is a polynomial setting the order $m \in \mathbb{N}$ of the kernel and $E_1$ is the exponential integral distribution. This particular function has a finite value at $\theta = 0$,
\begin{equation}
    g_m(0) = \frac{1}{2\pi}\left[\frac{\gamma}{2} - \ln\left(\sqrt{2}\epsilon\right) + Q_m(0)\right]
\end{equation}
where $\gamma = 0.5772156649$ is Euler's constant.

After replacing the green's function eq.~(\ref{eq:convol}) can be discretised and solved by multiplication in Fourier space. When numerically evaluating eq.~\ref{eq:convol} using a discrete Fourier transform (DFT), implicitly periodic boundary conditions are assumed. One can account for isolated boundary conditions by zero padding. The solution being to pad out $\kappa$ to twice its original size, assigning zero to all the newly created cells. One then defines $g_m^{(ij)}(\theta^{(ij)})$ on this new grid and calculates the components of the FT of the lensing potential as 
\begin{equation}
    \hat\psi^{(qp)} = 2\hat g_m^{(qp)}\hat{\kappa}^{(qp)},    
\end{equation}
where the hat symbolises the Fourier transform and the upper $(qp)$ index corresponds to the Fourier space position $\boldsymbol k^{(qp)}$. Finally we recover the lensing potential by inverting the FT and removing the padded area.

\paragraph*{Computation of the deflection angles} \ During this work we have identified two methods of computing quantities that are derivatives of the lensing potential, such as the deflection angles $\boldsymbol{\alpha}^{(ij)}$ which allow to calculate the coordinates in the source plane, $\boldsymbol{\beta}^{(ij)}$, where each grid point maps to. The simplest manner is to use finite differences. To second order the components of the deflection angles will be.
\begin{equation}
    \begin{cases}
    \alpha_{1}^{(ij)} = \frac{\psi^{(i+1,j)} - \psi^{(i-1,j)}}{2\Delta} \\
    \alpha_{2}^{(ij)} = \frac{\psi^{(i,j+1)} - \psi^{(i,j-1)}}{2\Delta}
    \end{cases}
\end{equation}
The second method is to use the differentiation property of the convolution, see Eq.~(\ref{eq:diff_prop}), such that we express the different quantities with respect to analytical derivatives of the Green's function. This gives the deflection angle components, 
\begin{equation}
    \alpha_i = \partial_i g_m * 2\kappa
\end{equation}
using 
\begin{align}
\begin{split}
    \partial_i g_m = -\frac{1}{2\pi} &\left\{  \frac{\theta_i}{\theta^2}\left[1 - \exp\left(-\frac{\theta^2}{2\epsilon^2}\right)\right] \right.\\ 
    &\left.+ \frac{\theta_i}{\epsilon^2}\exp\left(-\frac{\theta^2}{2\epsilon^2}\right)\left[Q_m\left(\frac{\theta}{\epsilon}\right) - \frac{\epsilon}{\theta}Q'_m\left(\frac{\theta}{\epsilon}\right)\right]\right\}
\end{split}
\label{eq:def_kernel}
\end{align}
for the derivative of the Green's function, with $\partial_i g_m|_{\theta = 0} = 0$. We refer to this method as the spectral method.

\paragraph*{Computation of the Distortion matrix} \ Similarly, one can define the components of the distortion matrix $\mathbfss{A}$ either through finite differences or using analytical derivatives,
\begin{equation}
    \mathbfss{A}_{ij} = \delta_{ij} - \partial_i\partial_j g_m * 2\kappa
\end{equation}
where $\delta_{ij}$ is the Kronecker delta symbol. For which we require two expressions for the different possible combinations of derivatives. Leading to,
\begin{align}
\begin{split}
    \partial_i^2 g_m =& -\frac{1}{2\pi}\left\{  \frac{(-1)^i(\theta_2^2 - \theta_1^2)}{\theta^4}\left[1 - \exp\left(-\frac{\theta^2}{2\epsilon^2}\right)\right]\right.\\
    &+ \frac{1}{\epsilon^2}\exp\left(-\frac{\theta^2}{2\epsilon^2}\right)\left( \left(1 - \frac{\theta_i}{\epsilon^2}\right)\left[Q_m\left(\frac{\theta}{\epsilon}\right) - \frac{\epsilon}{\theta}Q'_m\left(\frac{\theta}{\epsilon}\right)\right]\right.\\
    &\left.\left.\frac{\theta_i^2}{\theta^2}\left[\left(\frac{\theta^2+\epsilon^2}{\theta\epsilon}\right)Q_m'\left(\frac{\theta}{\epsilon}\right) - \frac{\epsilon}{\theta}Q''_m\left(\frac{\theta}{\epsilon}\right) - 1\right]\right)\right\}
\end{split}
\end{align}
for the diagonal terms, which evaluated at the origin yields $\partial_i^2 g_m|_{\theta=0} =  \frac{1}{2\pi\epsilon^2}\left(\frac{1}{2} + Q_m(0) - Q_m''(0)\right)$, and 
\begin{align}
\begin{split}
    \partial_i\partial_j g_m =& -\frac{1}{2\pi}\left\{\frac{\theta_i\theta_j}{\theta^4} \left[-\frac{\theta^2
    }{\epsilon^2} \exp\left(-\frac{\theta^2}{2\epsilon^2}\right) +2\left(1 - \exp\left(-\frac{\theta^2}{2\epsilon^2}\right)\right)\right]\right.\\
    & - \frac{\theta_i\theta_j}{\epsilon^2} \exp\left(-\frac{\theta^2}{2\epsilon^2}\right) \left( -\frac{1}{\epsilon^2} \left[Q_m\left(\frac{\theta}{\epsilon}\right) - \frac{\epsilon}{\theta} Q_m'\left(\frac{\theta}{\epsilon}\right)\right] \right.\\
    & \left.\left.+ \frac{1}{\theta^2} \left[\frac{\theta^2 + \epsilon^2}{\theta\epsilon}Q_m'\left(\frac{\theta}{\epsilon}\right) - Q_m''\left(\frac{\theta}{\epsilon}\right)\right]\right)\right\}
\end{split}
\end{align}
for the cross terms which yields $\partial_i\partial_j g_m|_{\theta=0} =  0$.

The convergence of both these schemes is discussed in App.~(\ref{app:convergence}) where one can see that the spectral scheme is overall more accurate and presents a faster convergence rate. On the basis of these tests we privilege the use of the spectral scheme.

\section{Numerical Convergence}
\label{app:convergence}

\begin{figure}
    \centering
    \includegraphics[width = \linewidth]{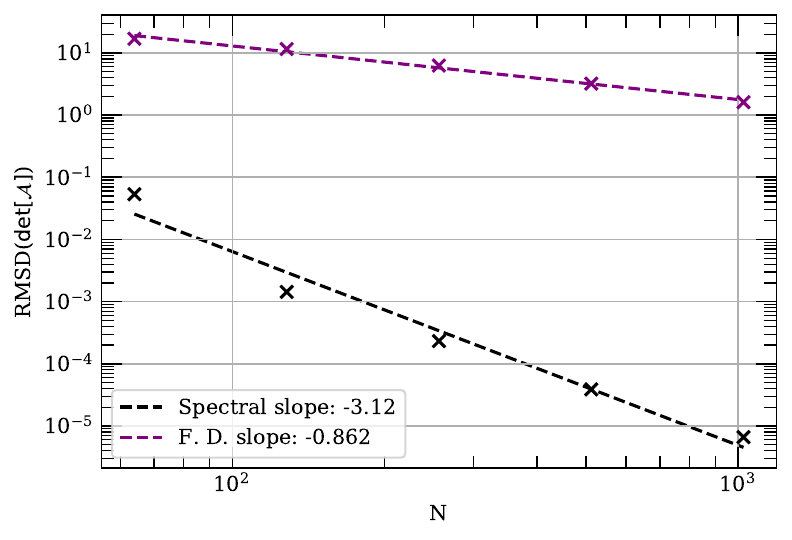}
    \caption{RMSD of the determinant of distortion matrix with respect to the analytical solution for increasing resolution. The black curve corresponds to the FD scheme and the purple curve corresponds to the spectral method. We observe that the spectral method is overall more accurate and converges at a faster rate than the FD scheme.}
    \label{fig:RMS}
\end{figure}

Here we test the convergence of the numerical implementations discussed in App.~(\ref{app:num_imp}). To do so we set up a simple test problem to which we can find an analytical solution, a Gaussian surface density field
\begin{equation}
    \kappa = \frac{K_m}{2\pi\sigma^2}\exp\left(-\frac{\theta^2}{2\sigma^2}\right)
\end{equation}
where $\theta =\sqrt{\theta_1^2 + \theta_2^2}$ is a radial coordinate, $K_m$ is the physical mass of the profile in units of the critical surface density and $\sigma$ is its width.
We solve Eq.~(\ref{eq:poisson}) to obtain an analytical solution for the lensing potential
\begin{equation}
 \psi = \frac{K_m}{4\pi}\left[ \log\left(\frac{\theta^4}{4\sigma^4}\right) -2 E_i\left(\frac{\theta^2}{2\sigma^2}\right)\right] + C 
\end{equation}
where $E_i$ is the exponential integral and $C$ is a constant gauge term. The main output of this code being the deflection angles and magnification it is of interest to study them directly. As we have a strong radial symmetry to our problem the solution is fully described by the norm $\alpha$ of the deflection angle.
\begin{equation}
    \alpha = \frac{K_m}{\pi\theta}\left[1 - \exp\left(-\frac{\theta^2}{2\sigma^2}\right)\right].
\end{equation}

We finally give the three independent components of the distortion tensor $\mathbfss{A}$ for $r<r_d$
\begin{equation}
    \begin{cases}
 \mathbfss{A}_{11} = 1 - \frac{K_m}{\pi}\left[\frac{\theta_2^2 - \theta_1^2}{r^4}  \left(1 - \exp\left(-\frac{\theta^2}{2\sigma^2}\right)\right) + \frac{\theta_1^2}{\theta^2\sigma^2}  \exp\left(-\frac{\theta^2}{2\sigma^2}\right)\right]\\
    \mathbfss{A}_{22} = 1 - \frac{K_m}{\pi}\left[\frac{\theta_1^2 - \theta_2^2}{\theta^4}  \left(1 - \exp\left(-\frac{\theta^2}{2\sigma^2}\right)\right) + \frac{\theta_2^2}{\theta^2\sigma^2} \exp \left(-\frac{\theta^2}{2\sigma^2}\right)\right]\\
    \mathbfss{A}_{12} = \frac{K_m}{\pi} \frac{\theta_1\theta_2}{\theta^4}  \left[2\left(1 - \exp\left(-\frac{\theta^2}{2\sigma^2}\right)\right) - \frac{\theta^2}{\sigma^2}\exp\left(-\frac{\theta^2}{2\sigma^2}\right)\right]
    \end{cases}
\end{equation}

With this solution we now compute the Root Mean Square Deviation (RMSD) between the numerical solution and the analytical solution, imposing in the numerical case that the total mass of the profile is conserved. We repeat this while increasing the resolution. The result of this is reproduced in Fig.~\ref{fig:RMS} along with a power law fit giving an indication of the convergence rate.

\section{Scale dependence} \label{app:scaledependence}

It should be ensured that the observed flux ratio anomalies are produced by effects that cannot be absorbed by the lens model, e.g. the displacement of images. We have already investigated this less extensively in Section \ref{sec:FR} by considering the flux-ratios induced by the full convergence field and the high-pass filtered field which only contained contributions from scales smaller than the Einstein radius. Here, we show in more detail, how the halo and non-halo contributions behave as a function of scale.

Drawing inspiration from \cite{Inoue2012}, we split the contribution of different scales using $k-$space filters,
\begin{equation}
    \hat{h}_{k_\text{s}} := \exp\left\{-\frac{(k - k_\text{s})^2}{2\sigma_k^2}\right\},
\end{equation}
to isolate the scale $k_\text{s}$. Here we have used a width of $\sigma_k \simeq 0.043\text{ arcsec}^{-1}$ and have selected the modes using a linear step, $\Delta k_\text{s} \simeq 0.142\text{ arcsec}^{-1}$, as such we have $\Delta k_\text{s} \simeq  3.28\sigma_k$. Repeating this process for multiple values of $k_\text{s}$, we measure flux ratio anomalies as well as the displacement of the images. From this information we generate a displacement spectrum, showing which scales impact image positions the most, and a flux ratio anomaly spectrum to identify the contribution of different scales to the total measured flux ratio anomalies. 

\begin{figure}
    \centering
    \includegraphics[width = 0.95\linewidth]{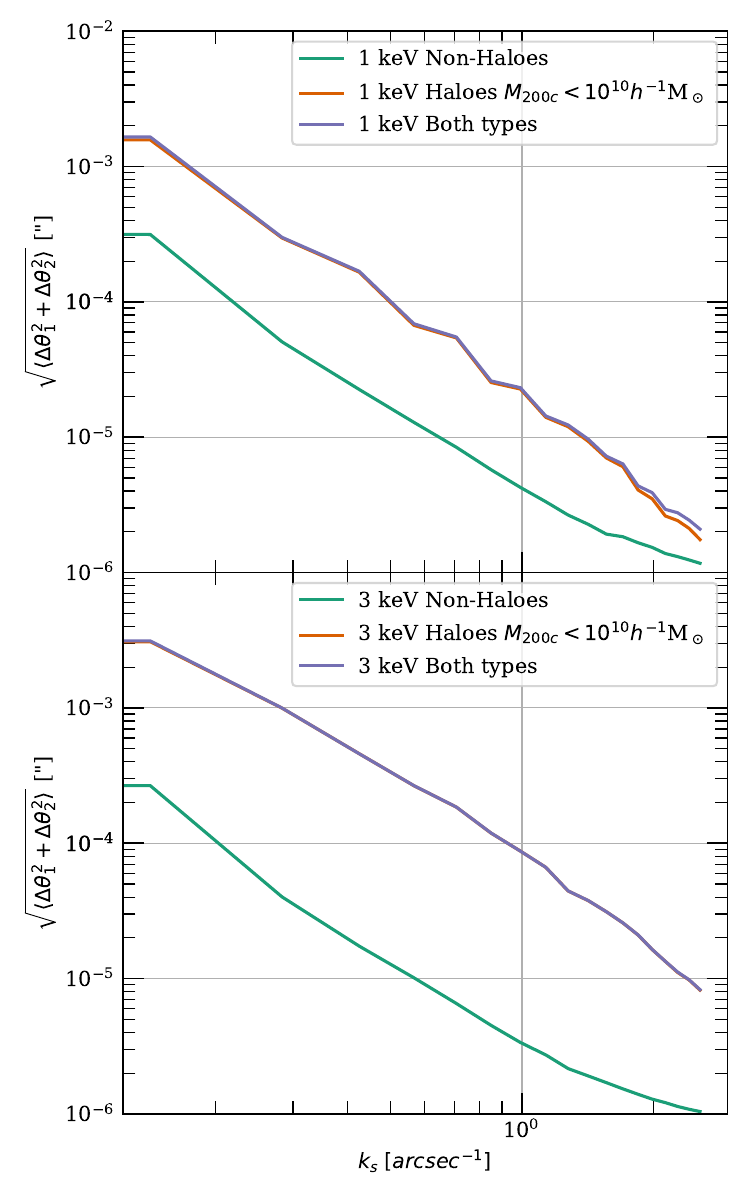}
    \caption{Scale dependent estimate of the RMS displacement of the images with respect to their unperturbed positions. Top: in a 1keV cosmology, center: in a 3keV cosmology and bottom: comparing CDM haloes of different masses. Haloes are far more efficient at displacing the images than Non-halo structures.}
    \label{fig:spec_displacements}
\end{figure}

In Figure \ref{fig:spec_displacements} we show the RMS displacements as a function of $k_s$. We can see that haloes are more efficient at moving the images than non-halo structures. We can also identify that for perturbations of below the scale of the Einstein radius, $k_\text{E} = 1/\theta_\text{E} \simeq 1.1$ arcsec$^{-1}$, the images are not significantly displaced in any scenario. Therefore, the observed signal would not be absorbed by lens model fitting techniques. 

\begin{figure}
    \centering
    \includegraphics[width = 0.95\linewidth]{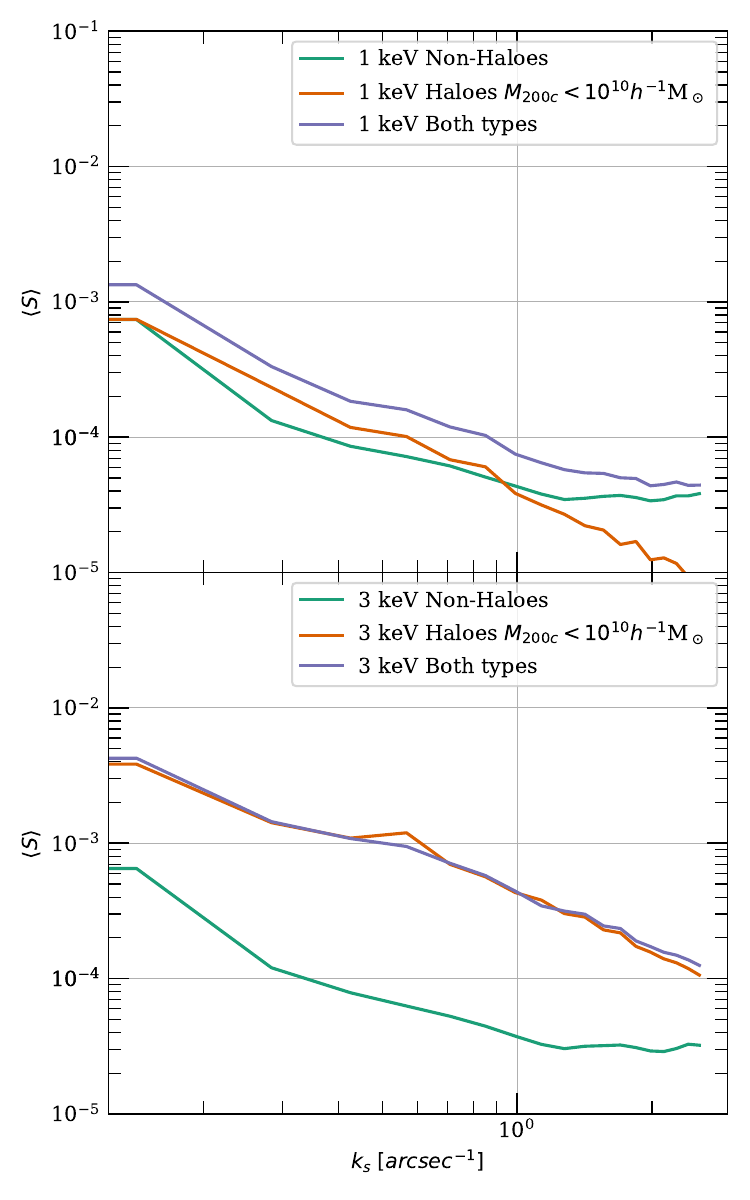}
    \caption{Scale dependent estimate of the observed mean flux ratio anomaly. Top: in a 1keV cosmology, center: in a 3keV cosmology and bottom: comparing CDM haloes of different masses. In the 1keV scenario Non-halo structures are a major component of the signal at all scales. In the 3 keV case they still represent $\sim$10 percent of the observed signal. The relative contributions of the different components are relatively scale-independent.}
    \label{fig:spec_anomalies}
\end{figure}

In Fig.~(\ref{fig:spec_anomalies}) we show  the mean flux ratio anomaly as a function of scale. In the 1 keV model Non-halo structures play a major role in the observed flux-ratio anomalies at all scales. However, when looking at the colder 3keV model we can see that while their contribution has not significantly declined, that of haloes has risen rapidly superseding them as the dominant contributor at all scales. None the less even in this model Non-halo structures still contribute at the order of 10 percent to the final measured signal. This seems to be the case no matter the filtering scale. We conclude that our relative measurements from Section \ref{sec:FR} are quite independent of the filtered scale. Therefore, our conclusions remain valid even for cases where an observational fitting procedure might absorb large scale contributions.

%%%%%%%%%%%%%%%%%%%%%%%%%%%%%%%%%%%%%%%%%%%%%%%%%%

% Don't change these lines
\bsp	% typesetting comment
\label{lastpage}
\end{document}